\documentclass[aps,pra,10pt,twocolumn,floatfix]{revtex4-2}

\usepackage[utf8]{inputenc}
\usepackage{times}
\usepackage{graphicx}
\usepackage{subfigure}
\usepackage{dcolumn}
\usepackage{bm}
\usepackage{xcolor}
\usepackage{amsmath,amssymb}
\usepackage{verbatim}
\usepackage{mathtools}
\usepackage{physics}
\usepackage{epstopdf}
\usepackage{bbold}
\usepackage{verbatim}

\usepackage{hyperref}
\hypersetup{
	colorlinks=true,
	linkcolor=blue,           
	citecolor=blue,           
	filecolor=magenta,        
	urlcolor=cyan
}

\begin{document}

\title{Quantum Equilibrium Propagation for efficient training of quantum systems based on Onsager reciprocity}

    \author{Clara C. Wanjura}
    \address{Max Planck Institute for the Science of Light, Staudtstraße 2, 91058 Erlangen, Germany}
	
    \author{Florian Marquardt}
    \address{Max Planck Institute for the Science of Light, Staudtstraße 2, 91058 Erlangen, Germany}
    \address{Department of Physics, University of Erlangen-Nuremberg, 91058 Erlangen, Germany}
	
	\date{\today}


\begin{abstract}
    The widespread adoption of machine learning and artificial intelligence in all branches of science and technology has created a need for energy-efficient, alternative hardware platforms. While such neuromorphic approaches have been proposed and realised for a wide range of platforms, physically extracting the gradients required for training remains challenging as generic approaches only exist in certain cases. Equilibrium propagation (EP) is such a procedure that has been introduced and applied to classical energy-based models which relax to an equilibrium. Here, we show a direct connection between EP and Onsager reciprocity and exploit this to derive a quantum version of EP. This can be used to optimize loss functions that depend on the expectation values of observables of an arbitrary quantum system.
    Specifically, we illustrate this new concept with supervised and unsupervised learning examples in which the input or the solvable task is of quantum mechanical nature, e.g., the recognition of quantum many-body ground states, quantum phase exploration, sensing and phase boundary exploration.
    We propose that in the future quantum EP may be used to solve tasks such as quantum phase discovery with a quantum simulator even for Hamiltonians which are numerically hard to simulate or even partially unknown.
    Our scheme is relevant for a variety of quantum simulation platforms such as ion chains, superconducting qubit arrays, neutral atom Rydberg tweezer arrays and strongly interacting atoms in optical lattices.
\end{abstract}

\maketitle

\section{Introduction}

As deep learning and artificial intelligence are adopted in all braches of science and technology, the increasing complexity of neural networks has led to an exponential increase in energy consumption and training costs. This has created a need for more efficient alternatives, sparking the rapidly developing field of neuromorphic computing~\cite{markovic2020physics}, which explores a variety of different platforms~\cite{wetzstein2020inference,shastri2021photonics} to design physical, analogue neural networks.

Existing backpropagation-based training strategies for neuromorphic platforms include in-silico training, requiring a faithful digital model of the system, and physics-aware backpropagation~\cite{wright2022deep}, combining physical inference with a simulated backward pass which relaxes these constraints. However, it is a central question whether not only inference but also training can exploit the physical dynamics~\cite{momeni2024training}, making full use of the energy efficiency of neuromorphic systems. For example, feedback-based parameter shifting 
does not require any simulation 
but scales unfavourably with the network size~\cite{bartunov2018assessing}.
Moving towards physical implementations of efficient backpropagation,  strategies for specific types of non-linearities have been developed~\cite{psaltis1990holography, guo2021backpropagation,spall2023training}, as well as approaches performing backpropagation only on the linear components~\cite{hughes2018training,pai2023experimentally}.
Another novel recent approach enables ``forward-forward'' type gradient calculation in systems which perform sequential information processing~\cite{Momeni2023Backpropagation}. 
Furthermore, efficient measurements of gradients via scattering experiments can be performed in optical systems that employ a framework recently developed to produce nonlinear computation with linear wave setups~\cite{wanjura2023fully}--- such nonlinear processing was also recently demonstrated in Refs.~\cite{yildirim2023nonlinear,xia2023deep,Momeni2023Backpropagation}.

\begin{figure}[htbp]
    \centering
    \includegraphics[width=.49\textwidth]{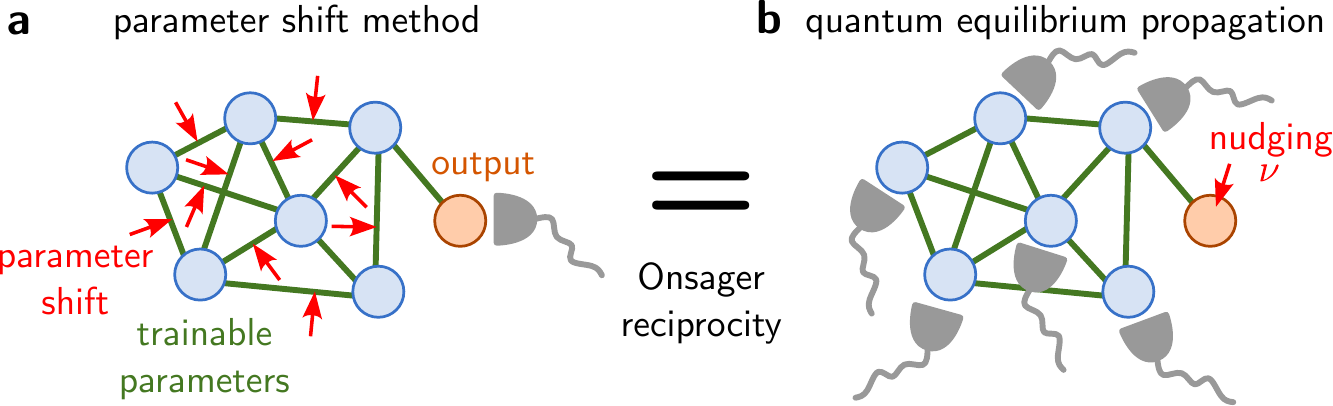}
    \caption{\textbf{The concept of quantum equilibrium propagation.} The goal is to efficiently and in a physical way obtain the gradient of some loss function (depending on expectation values measured at the ``output'' degrees of freedom of a quantum system) with respect to tuneable parameters.
    \textbf{a}~Rather than shifting $N$ parameters separately and measuring the output expectation value for each shift (parameter-shift method), Onsager reciprocity dictates that the same information can be extracted by \textbf{b}~shifting, i.e. nudging, only the parameters coupling to the output observables and (in a single go) measuring the response of all $N$ operators coupled to the training parameters (quantum equilibrium propagation). The latter procedure is more efficient as it requires only a single response experiment (or at most a small number of order 1, when some non-commuting observables have to be measured), whereas the parameter shift method requires a number of experiments scaling linearly with the number of parameters.}
    \label{fig:QEPConcept}
\end{figure}%

General approaches for physical backpropagation so far only exist in two classes of physical systems: Hamiltonian Echo Backpropagation~\cite{lopez2021self}, which applies to essentially lossless systems in which a time-reversal operation can be implemented, 
and
equilibrium propagation (EP)~\cite{scellier2017equilibrium,scellier2021deep}, which applies to energy-based, equilibrating systems.

EP stands in the tradition of contrastive learning approaches 
comparing measurements obtained from two different equilibria and using feedback to update parameters.
Concretely, EP consists of two phases: the free and the nudged phase. In the free phase, the input is fixed
and the system relaxes into its equilibrium state. In the nudged phase, the output is ``nudged'' closer towards the target output
and the system evolves to a new equilibrium. Comparing the system state in the free and the nudged phase, respectively, one obtains the necessary gradients which are then used to update parameters.

Since its introduction in 2017, EP has been investigated thoroughly~\cite{kendall2020training,martin2021eqspike,ernoult2020equilibrium,scellier2022agnostic,falk2023contrastive,wang2024training} and a variant, coupled learning, has been developed~\cite{stern2021supervised,stern2023physical}. In particular, EP was proposed for training nonlinear resistor circuits~\cite{kendall2020training}, systems of coupled phase oscillators~\cite{wang2024training}, was further adapted to spiking networks~\cite{martin2021eqspike}, to implement continual parameter updates~\cite{ernoult2020equilibrium} and a dynamical version was developed~\cite{scellier2022agnostic}.
Experimentally, EP has been applied to train electronic systems~\cite{dillavou2022demonstration,dillavou2023machine}, elastic networks~\cite{altman2023experimental} and even a memristor crossbar array~\cite{oh2023memristor}.
Furthermore, a classical Ising model has been trained using a quantum annealer to efficiently reach equilibrium~\cite{laydevant2023training}.

Given the elegance of the EP approach, it is a natural question to ask whether the procedure can be extended to quantum systems to train a fully quantum Hamiltonian via a nudging procedure similar to classical EP.
Indeed, we will show that there is a direct connection between EP and Onsager reciprocity and exploit it to derive a quantum version of EP (QEP). This new approach can be used to train efficiently arbitrary quantum systems, including specifically the highly tuneable quantum many-body systems realized nowadays in quantum simulators~\cite{altman2021quantum}.

We illustrate this new concept with supervised and unsupervised learning examples of a quantum-mechanical nature. Specifically, we investigate the recognition of quantum many-body phases and introduce as new concepts the exploration of phase diagrams with quantum simulator platforms via efficient gradient descent optimization enabled by QEP as well as the optimisation of sensitivity (e.g. for sensing applications) and phase boundary exploration.
QEP is applicable to systems which are hard or impossible to simulate classically and can be employed even in settings in which the Hamiltonian is only partially known or partially accessible.

In terms of the general question of using quantum devices for learning tasks, the area of quantum machine learning~\cite{schuld2015introduction,biamonte2017quantum}
by now has a long history. There have been some ideas of how to learn to reproduce quantum states by adapting tuneable parameters (``quantum Boltzmann machines'', see \cite{biamonte2017quantum}). The major research efforts in this domain are, however, spent on variational quantum circuits, which require digital quantum computing platforms for implementation (possibly even fault-tolerant), going significantly beyond the resources that we are going to assume here.

\section{Quantum equilibrium propagation}
\subsection{Onsager reciprocity}
Consider a parameterized Hamiltonian 
\begin{equation}
    {\hat H}(\lambda) = \sum_j \lambda_j {\hat A}_j
\end{equation}
and its ground state $\left| \Psi(\lambda) \right\rangle$. A small static force coupling to ${\hat A}_j$ (entering as a term $\delta \lambda_j {\hat A}_j$ inside ${\hat H}$) will produce a linear response in the expectation value $\left\langle {\hat A}_\ell \right\rangle$, given by
\begin{equation}
\chi_{\ell j} = {\partial \over \partial \lambda_j} \left\langle \Psi(\lambda) \right| {\hat A}_\ell \left| \Psi(\lambda) \right\rangle .
\end{equation}
Onsager reciprocity guarantees the symmetry of the susceptibility $\chi_{j\ell}=\chi_{\ell j}$, i.e. the same effect will be produced by a force acting on ${\hat A}_\ell$ influencing the expectation value $\left\langle {\hat A}_j \right\rangle$, see Fig.~\ref{fig:QEPConcept}, i.e.,
\begin{align}\label{eq:Onsager}
    \frac{\partial}{\partial \lambda_j} \langle \hat A_\ell\rangle & = \frac{\partial}{\partial \lambda_\ell} \langle \hat A_j\rangle.
\end{align}
Onsager reciprocity~\cite{Onsager1931Reciprocal} can be derived in many ways, also for the quantum case~\cite{Kubo1966The}. However, the most elementary approach for static situations such as the one considered here and applied to the ground state in particular uses first-order perturbation theory for the deformation of the ground state, $\partial_{\lambda_j} \ket{\Psi(\lambda)} = (E(\lambda) - H(\lambda))^{-1} (A_j - \langle A_j\rangle)\ket{\Psi(\lambda)}$.
Expression~\eqref{eq:Onsager} also holds for thermal states for which the expectation values above are replaced by $\langle \hat A_j\rangle = \Tr(\hat \rho \hat A_j)$, so the following results apply for arbitrary-temperature quantum equilibrium states.
In the case of ground-state degeneracy, expression~\eqref{eq:Onsager} also holds with $\hat \rho = \sum_j \ketbra{\Psi_j(\lambda)}{\Psi_j(\lambda)}$ in which $j$ sums over all degenerate states; this is approximately equivalent to a thermal state at small, but non-vanishing, temperature.
Classical Onsager reciprocity is equivalent to what has been termed the ``Fundamental Lemma'' for classical EP~\cite{scellier2017equilibrium,scellier2021deep}.

\subsection{Onsager reciprocity as a basis for quantum equilibrium propagation}
We will now show that this well-known result~\eqref{eq:Onsager} gives us access to a general version of equilibrium propagation for quantum systems. To understand that, we first consider supervised learning. Let us assume that the set of operators ${\hat A}_j$ is split into degrees of freedom relating to the input $j\in {\mathcal S}_{\rm in}$, trainable variables $j\in {\mathcal S}_{\rm train}$, and the output $j\in {\mathcal S}_{\rm out}$.
Accordingly, the set of parameters $\lambda$ is split into the input $x$ containing the parameters corresponding to the input $\lambda_j = x_j$,
the set of training parameters $\theta$ with $\lambda_j = \theta_j$
and the set of couplings to the output observables $\nu$ with $\lambda_j = \nu_j$.
Hence, a general QEP Hamiltonian is of the form $\hat H(x, \theta, \nu)$ in which $\nu = 0$ during inference.
For any given training sample, the input $x$ is fixed by applying a field to all input degrees of freedom (we may write $\lambda_k = x_k$ for $k\in {\mathcal S}_{\rm in}$, with $x$ representing the input vector). The output is then read off as the expectation values $y_\ell = \left\langle {\hat A}_\ell \right\rangle$ in the operators $\ell\in {\mathcal S}_{\rm out}$. Note that ${\hat A}_\ell$ can be chosen as a projector, in which case $y_\ell$ becomes the probability of obtaining a particular outcome in a measurement; this is useful for classification tasks.

In supervised learning, we are interested in adjusting the trainable parameters $\theta$ in order to ``nudge'' the output closer to the desired target output $y^{\rm target}(x)$, for any given input $x$. More generally, we aim to reduce the loss function ${\mathcal L}(y,y^{\rm target})$, or rather its average $\bar {\mathcal L}$ over many training samples $(x,y^{\rm target}(x))$,  via gradient descent: $\delta \theta = - \eta {\partial {\bar {\mathcal L}} / \partial \theta_j}$. To do this, we need to obtain the influence of a change in $\theta_j$ on any of the outputs $y_\ell$. For a given fixed training sample, this is just the susceptibility $\chi_{\ell j}(\lambda)=\partial \left\langle {\hat A}_\ell \right\rangle / \partial \theta_j$.
Evaluating $\chi_{\ell j}$ for all possible trainable parameters $j$, see Fig.~\ref{fig:QEPConcept}~\textbf{a},  scales unfavourably, requiring a number of different experiments that scales linearly in the number $N$ of these parameters. Accessing the training gradient in this way amounts to the parameter-shift method which is always applicable in any neuromorphic platform but should generally be avoided whenever possible due to this unfavourable scaling. However, Onsager reciprocity, Eq.~\eqref{eq:Onsager}, tells us that we can also access the susceptibility $\chi_{\ell j}$ by performing an alternative, much more efficient experiment that instead reveals $\chi_{j\ell}$: apply a small force $\nu$  acting on the outputs and observe its influence on the expectation values of the degrees of freedom $\left\langle {\hat A}_k \right\rangle$  connected to the trainable parameters: $\chi_{j\ell}(\lambda)=\partial \left\langle {\hat A}_j \right\rangle / \partial \nu_\ell$, see Fig.~\ref{fig:QEPConcept}~\textbf{b}. This would seem to require a number of experiments that scales with the number of outputs $N_{\rm out}$, already typically much smaller than the number of trainable parameters. However, by evaluating explicitly the desired gradient of the loss function, it becomes apparent that only a single experiment is in fact needed: in this experiment, a force vector $\nu = \partial {\mathcal L} / \partial y$, the so-called error signal, is applied to the output degrees of freedom. 

In this way, Onsager reciprocity teaches us how to translate the classical equilibrium propagation approach to quantum devices. Quantum Hamiltonians of arbitrary structure can be considered.

\subsection{QEP procedure}

We now explicitly summarize the QEP procedure. For clarity, we will from now on denote the output observables by $\hat O_\ell$.  For a gradient-descent parameter update, we need to compute the derivative of the loss function,
\begin{align}
    \frac{\partial}{\partial \theta_j} \mathcal{L}(y,y^\mathrm{target}(x))
    & = \sum_\ell \frac{\partial \mathcal{L}}{\partial y_\ell} \frac{\partial y_\ell}{\partial \theta_j} = \varepsilon \frac{\partial y}{\partial \theta_j} 
\end{align}
in which the error signal  vector has components $\varepsilon_\ell = {\partial \mathcal{L}}/ {\partial y_\ell}$. For a mean-square-error loss function, we would have $\varepsilon=2 (y - y^\mathrm{target}(x))$.

The QEP procedure for supervised learning can be summarized as follows.
(i)~Free phase: The nudging forces are off ($\nu_j=0$ for all $j$) and the output expectation values $y_\ell = \langle \hat O_\ell\rangle$ as well as the expectation values of all operators associated with trainable parameters $\hat A_j$ are measured in the ground state of the Hamiltonian $\hat H(x,\theta,0)$.
(ii)~Nudged phase:
We compute the error signal $\varepsilon$ and use it to nudge the Hamiltonian $\hat H(x,\theta,\nu = \beta \varepsilon)$ by switching on the couplings to the output observables, adding a term $\sum_\ell \nu_\ell {\hat A}_\ell$ to the Hamiltonian. The couplings are given by the vector  $\nu = \beta \varepsilon$, in which $\beta$ is a small parameter (keeping with the notation for classical EP~\cite{scellier2017equilibrium}; this is unrelated to the inverse temperature).
We again measure the expectation values of all observables $\hat A_j$.
(iii)~Parameter update: Using Onsager reciprocity, Eq.~\eqref{eq:Onsager}, ${\partial} \langle \hat O_\ell\rangle / {\partial \theta_j}= {\partial} \langle \hat A_j\rangle / {\partial \nu_\ell} $, we can approximate the gradient ${\partial} \langle \hat O_\ell\rangle / {\partial \theta_j}$ 
and hence arrive at:
\begin{align}\label{eq:QEPDerivative}
    \frac{\partial}{\partial \theta_j} \mathcal{L}(y,y^\mathrm{target}(x))
    &
    \approx \frac{\langle \hat A_j\rangle\big\rvert_{\nu = \beta \varepsilon} - \langle \hat A_j\rangle\big\rvert_{\nu = 0}}{\beta}.
\end{align}

In a similar spirit as for classical Equilibrium Propagation, one may consider variants which combine positive and negative nudging~\cite{scellier2023energybased}, i.e., approximate the gradient using $({\langle \hat A_j\rangle\big\rvert_{\nu = \beta \varepsilon} - \langle \hat A_j\rangle)\big\rvert_{\nu = -\beta \varepsilon}} / {(2 \beta)}$, which empirically performs better for finite nudging.

\subsection{Practical requirements}
We now discuss the most important practical considerations for implementing quantum equilibrium propagation (QEP) in any experimental platform.

Above all, the platform needs to have tuneable couplings $\theta$, whose number preferably should be easy to scale up with growing system size. Such tuneable couplings have been developed for many quantum simulators \cite{altman2021quantum} and quantum computing platforms  by now. Examples include: (i) ion chains, where spin-spin couplings can be mediated and engineered via the vibrational modes of the chain, employing suitable Raman transitions., e.g. \cite{kim2009entanglement}. (ii) superconducting-qubit arrays, as employed for quantum computing, with current-tuneable couplers between neighboring qubits \cite{chen2014qubit}. (iii) neutral-atom Rydberg tweezer arrays providing tuneable spin-spin couplings \cite{steinert2023spatially}. (iv) strongly interacting atoms in optical lattices with spatially engineered hopping and interactions, e.g. based on holographic potential shaping \cite{bakr2009quantum}. Other platforms, e.g. in optomechanical arrays, coupled microwave cavities, or coupled laser arrays, also demonstrate interesting tuneable coupling schemes, but they often operate out of equilibrium and are therefore not directly suitable for QEP, unless one can map them back to an equilibrium situation.

Beyond this primary requirement of a scalable number of tuneable couplings, QEP platforms also need ways to apply the output forces $\nu$ . This demands local fields, e.g. effective magnetic fields or qubit detunings, easily available in most platforms that are flexible enough to support tuneable couplings. In addition, the expectation values of both the output operators and of the coupling operators connected to trainable degrees of freedom should be measurable. Regarding the couplings, we note that the expectation values of interaction terms of the form $\hat X_j \hat X_k$ or similar can easily be measured even by observing the spin operators $\hat X_j$ individually (and multiplying outcomes). A Heisenberg-type coupling operator $\sum_{\alpha = x,y,z}\hat \alpha_j \hat \alpha_k$ would need three separate measurements, for the x,y,z components, performed in independent shots of the experiment, eventually obtaining the expectation value of the complete operator. Alternatively, one could carry out a collective (two-qubit) measurement. The latter is typically performed via an ancilla, as demonstrated for syndrome extraction in quantum error correction schemes, and therefore requires more experimental effort. We note that the statistical nature of quantum physics in any case requires many runs of the experiment to measure the expectation values of output variables (needed for inference) and of coupling operators (needed for training). Each of these runs involves an equilibration step.

Finally, QEP requires efficient means to approach the equilibrium state, i.e., for the zero-temperature limit, the ground state $\left| \Psi(\lambda) \right\rangle$ of the Hamiltonian. Efficient ground-state preparation of complex quantum many-body Hamiltonians is one of the most intensively researched questions in quantum simulation and quantum computing. For the purposes of QEP, we distinguish between two options: (i) hybrid approaches, where an external digital computer is employed during equilibration, and (ii) purely autonomous schemes. Hybrid approaches could rely on variational quantum eigensolvers \cite{tilly2022variational}, where an ansatz quantum circuit with continuously parametrized unitary gates is performed, the expectation value of the Hamiltonian is measured, and a classical optimization is performed to adapt parameters of the circuit. At first glance, the use of a classical optimizer to find the quantum ground state in this way might seem to contradict the basic premise of QEP or neuromorphic computing in general, i.e. using a physical system to do information processing. However, if the problem setting makes efficient use of the resulting quantum many-body ground state, even such a hybrid approach may still yield an advantage over an entirely classical digital device, in the same way that variational quantum eigensolvers are thought to be beneficial under the right circumstances as compared to numerical ground state search by classical algorithms. 

Purely autonomous equilibration schemes get rid of any feedback loop. In principle, coupling to a cold environment is sufficient, but recently there has been active research into speeding up equilibration. The techniques put forward often rely in one way or another on variations of quantum reservoir engineering\cite{poyatos1996quantum}, where dissipation is introduced deliberately, and which have, e.g., been used to stabilize quantum many-body states in superconducting circuits~\cite{ma2019dissipatively}. This can happen in the form of suitable continuous driving (inspired by laser cooling), e.g.~\cite{raghunandan2020initialization}, or else in the form of quantum circuits that introduce gates coupling the quantum many-body system to ancilla qubits which then may be periodically reset (``digital quantum cooling'', e.g.~\cite{polla2021quantum}) or other schemes to reduce entropy,
such as via suitable measurements~\cite{cotler2019quantum}. 

\section{Applications}
\subsection{Supervised learning: phase detection in a quantum many-body system}

The input considered above is classical, where QEP could be used to train a quantum device to perform an essentially classical machine learning task. However, we now turn to an important class of applications where a QEP-trained system can effectively receive input that is quantum instead of classical.
    
In general, this setting can be realized by starting from a Hamiltonian ${\hat H}_0(x)$, whose quantum ground state we want to analyze with the help of a QEP setup. By tuning the classical parameters $x$ (typically a few) we are able to realize different phases with different ground states, e.g., sweeping through some phase diagram in the case of a quantum many-body system. One task could consist in predicting, for any given $x$, the distinct quantum phase that the system assumes, possibly after seeing a few labeled training examples at a few parameter locations $x$. Another task could consist in approximating the entanglement entropy between some subsystems of ${\hat H}_0$ or predicting any other quantity of interest that can be derived in principle by inspecting the ground state but may be hard to extract directly by simple measurements.

\begin{figure}
    \centering
    \includegraphics[width=\columnwidth]{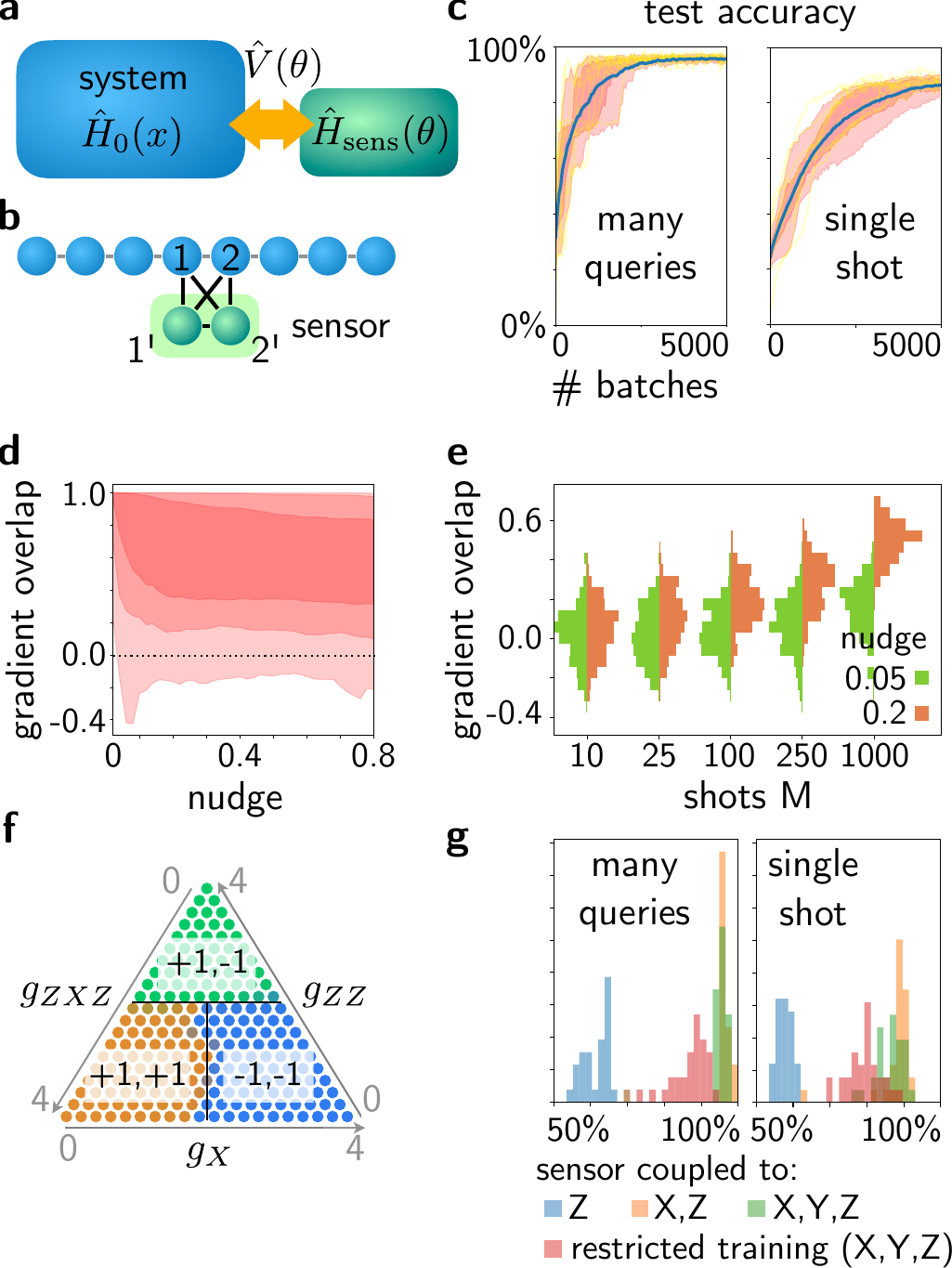}
    \caption{\textbf{Supervised learning: Learning recognition of quantum phases using QEP.} (a) Schematic of a trainable quantum sensor coupled to a system. (b) Specific example of a two-qubit sensor coupled to a 1D cluster transverse Ising Hamiltonian at two locations, where readout of ${\hat Z}_{1'}$ and ${\hat Z}_{2'}$ is supposed to indicate the phase. The 51 tuneable couplings are learned using QEP.  (c) Evolution of test accuracy during supervised training on the whole phase diagram (for a chain of length $N=8$). Multiple training runs (yellow), confidence intervals as areas (red; at 50\% and 80\%), and average accuracy (blue). Accuracies for ``many queries'' (asking whether the maximum-probability detector outcome matches correct phase) and ``single shot'' (probability to indicate correct phase in single quantum measurement) [batches of 10 training samples; projection noise for $M=10$ measurement shots per sample is accounted for; nudge parameter $\beta=0.4$]. (d) Overlap of batch-averaged gradient estimate with true gradient direction, vs. nudge parameter. Confidence intervals (red, 95\%, 80\%, 50\%) depict distribution over many batches  (batch size 10, no measurement shot noise). (e) Gradient overlap histograms vs. measurement shots $M$, for two different nudge parameters (batch size 10). (f) Test of phase recognition: probabilities of measuring the trained sensor in one of the three different combinations of $Z_{1'},Z_{2'}$ shown in orange/green/blue; true phase boundaries in black ($g_{\rm Z}+g_{\rm XX}+g_{\rm ZXZ}=4$). (g)  Histogram of final test accuracies for repeated training runs (parameters as above), for a sensor that only couples to ${\hat Z}$ operators in the chain (or only to ${\hat X}$ and ${\hat Z}$), and for a sensor trained only on a small patch in the middle of each phase but tested throughout (``restricted'').}
    \label{fig:QEP_phase_recognition}
\end{figure}

Any of these tasks can be addressed via QEP in the following way. We couple the system of interest, described by ${\hat H}_0(x)$, to a trainable physical sensor, described by ${\hat H}_{\rm sens}(\theta)$, Fig.~\ref{fig:QEP_phase_recognition}~\textbf{a}. Overall, coming back to our previous definitions, we thus have the full QEP Hamiltonian ${\hat H}(x,\theta)={\hat H}_0(x)+{\hat V}(\theta)+{\hat H}_{\rm sens}(\theta)$, where we assume that the couplings between the two systems reside inside ${\hat V}$ (see Fig.~\ref{fig:QEP_phase_recognition}). 

Ideally, the couplings inside the system of interest, ${\hat H}_0$, should be stronger than the couplings to the sensor and within that sensor. This will ensure that the system of interest is only weakly perturbed, while the recognition model can still react strongly to the features of the ground state $\left| \psi_0 \right\rangle $ of ${\hat H}_0$. However, in our numerical experiments (below) we have seen that even outside this scenario successful learning is possible.

Distinguishing quantum phases can serve as an important application, and it has been considered before in the context of quantum machine learning based on gate-based quantum computers, where a unitary circuit acts on a given ground state encoded in a multi-qubit register \cite{cong2019quantum,herrmann2022realizing,liu2023model}. In contrast to that, QEP relies on equilibration and moreover is far more general in the choice of systems – e.g. it does not require qubits as degrees of freedom nor the ability to perform gates nor any detailed knowledge of all aspects of the Hamiltonian. 

Both for this task as well as others analyzed below, we choose the cluster Ising Hamiltonian (see e.g.~\cite{liu2023model}) as an illustrative quantum many body system, 
\begin{align}\label{eq:ClusterIsing}
    {\hat H}_0=g_{\rm ZXZ} \sum_j {\hat Z}_{j-1} {\hat X}_j {\hat Z}_{j+1} - g_{\rm ZZ} \sum_j {\hat Z}_j {\hat Z}_{j+1} - g_{\rm X} \sum_j {\hat X}_j,
\end{align}
in which  ${\hat X}, {\hat Y}, {\hat Z}$ are the Pauli matrices. This model has three phases, including a topologically nontrivial one.  We will now regard  $x=(g_{\rm ZXZ}, g_{\rm ZZ}, g_{\rm X})$ as the input parameters. We add a sensor made of qubits that are coupled in all possible ways among each other (2-local, with terms like ${\hat X}_{\alpha} {\hat Y}_{\beta}$ or $Z_{\alpha}$ etc) and to a limited region in the chain (couplings ${\hat Z}_{\alpha} {\hat X}_j$ etc, where $j$ is a spin in the chain). The sensor qubits are measured in the $Z$-basis, and the resulting configuration is supposed to announce the detected phase. Here we used a mean-square-error loss function, although a categorical cross-entropy would have been suitable as well. We find that even a small-scale sensor of only two qubits, coupled to two spins in the chain, has sufficient expressivity to properly learn the known phase diagram of the cluster Ising model (Fig.~\ref{fig:QEP_phase_recognition}~\textbf{b},\textbf{c}), when trained in a supervised fashion using QEP. Inspecting the solution also reveals a surprise: the sensor-system couplings are not weak, and the spin configuration nearby the sensor is visibly perturbed, without however deteriorating the functionality. We also confirm that the quantum nature of the coupling is important. This can be ascertained by comparing to a sensor that is only allowed to couple to ${\hat Z}$ operators in the chain, which performs much more poorly (Fig.~\ref{fig:QEP_phase_recognition}~\textbf{g}). By contrast, allowing coupling to non-commuting obervables yields the observed good accuracy.

An important general aspect of QEP training is the unavoidable projection shot noise encountered in any quantum experiment, where expectation values are obtained by repeated measurements with individually discrete outcomes – in contrast to the classical situation. This hampers training, since the estimate of the gradient is noisy, with fluctuations $\sim 1/\sqrt{M}$ , where each of the $M$ shots requires a renewed equilibration. Fortunately, we find that this effect can be counteracted by employing an increased finite nudging strength $\mu$ , effectively boosting the contrast in estimating the response of expectation values. Finite nudging, however, leads to a deviation from the linear response that would yield the ideal gradients. Therefore ultimately there is a sweet spot, balancing shot noise vs. nonlinearity of the response, to obtain optimized training convergence (Fig.~\ref{fig:QEP_phase_recognition}~\textbf{d},\textbf{e}). In our numerical experiments, we find that the goal of minimizing the total number of experimental runs (i.e. number of training samples multiplied by number of shots per sample) is best achieved by keeping $M$ small and simply taking more batches. Apparently, this leads to more variety in the observed training data and better training performance. In experiments, the number of shots $M$ could also be reduced by coupling multiple sensors at different locations to the system, having them share their coupling values. Finally, in principle, there is an alternative to averaging over projective measurements, namely performing weak continuous measurements of the expectation values (a weakly coupled sensor is able to realize this). This could be performed without rethermalizing to the ground state.

\subsection{Unsupervised learning}
\begin{figure*}
    \centering
    \includegraphics[width=\textwidth]{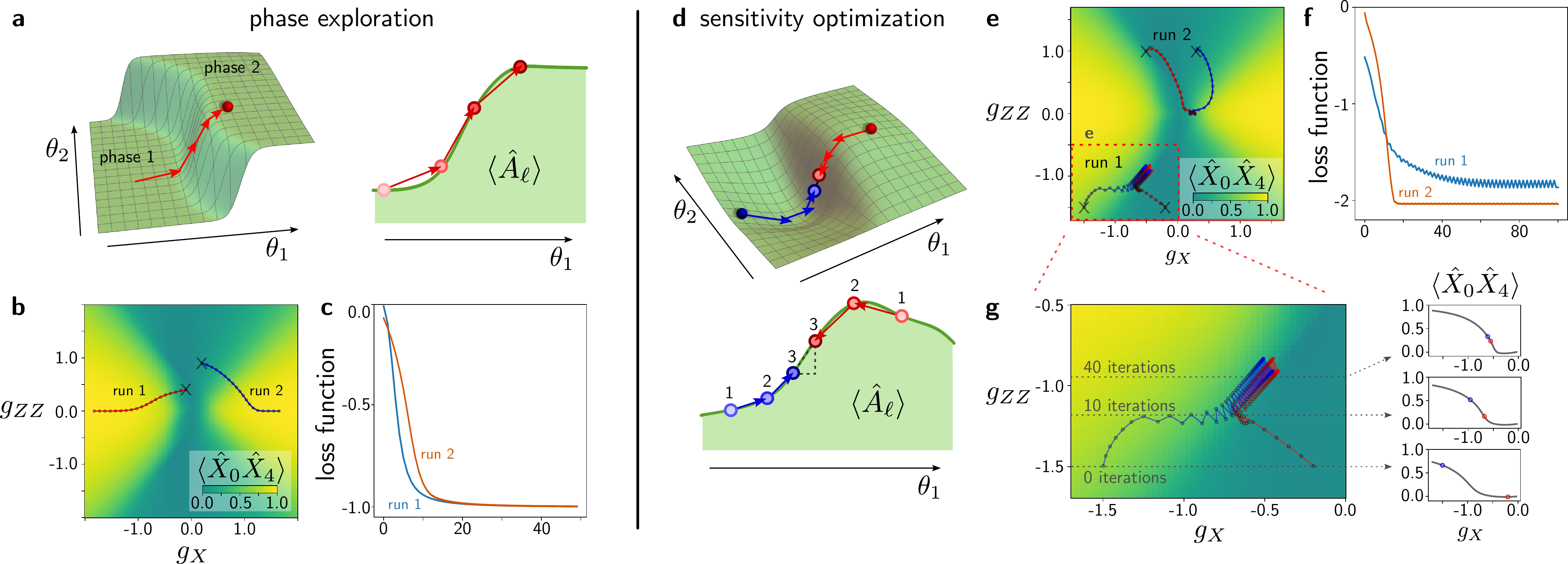}
    \caption{\textbf{Unsupervised learning applications: phase exploration and sensitivity optimization for sensing.}
   \textbf{a}~Sketch of phase space exploration using QEP. An expectation value of interest, $\langle \hat A_\ell\rangle$, is maximised.
   \textbf{b}~Two example trajectories in the phase diagram of the cluster Ising Hamiltonian~\eqref{eq:ClusterIsing} each starting at the position in the phase diagram marked with $x$ and \textbf{c}~the corresponding loss functions. The colours in \textbf{b} show $\langle {\hat X}_0 {\hat X}_4\rangle$.
   \textbf{d}~Sketch of the concept of sensitivity optimization. The derivative of a an observable of interest, $\langle {\hat A}_\ell\rangle$, w.r.t. a certain parameter (or w.r.t. a vector of parameters) is optimised. Concretely, this scheme can be used to devise optimal sensors or to find phase boundaries in a phase diagram.
   \textbf{e}-\textbf{g}~Optimization of $\partial\langle {\hat X}_0 {\hat X}_4\rangle/\partial g_X$ (magnetic field sensor).
    \textbf{e}~Two example runs, each showing the trajectories of two points in the phase diagram of the cluster Ising Hamiltonian~\eqref{eq:ClusterIsing} and \textbf{f}~the corresponding loss functions.
    The two points are inialized at the position marked with $\times$ and converge towards the phase boundary where the derivative w.r.t. $g_X$ is maximal. In the second run, \textbf{g}~the trajectories first converge towards a phase boundary and then follow it, suggesting that the technique may be employed to trace out phase boundaries.
    The plots on the right show cuts through the phase diagram at various steps during the training.
   }
    \label{fig:PhaseExploration}
\end{figure*}
    We now discuss two applications for unsupervised learning for which the gradients are used for optimization tasks.
    \paragraph{Phase exploration:}
    In the first example, we would like to explore the phase diagram of a quantum many-body system. In practice, this could be an interesting task for characterizing the capabilities of a quantum simulator when we would like to explore the phase diagram of a system which is partially unknown (e.g. some Hamiltonian terms are not known or cannot be tuned) and computationally hard to simulate classically. We can nevertheless ask whether phases exist that maximize (or minimize) certain expectation values and use QEP to explore the phase diagram and find such regimes.
    In this scenario, the relevant QEP-Hamiltonian of this unsupervised learning task is $\hat H(\theta, \nu)$, not containing any inputs $x$.

    Fig.~\ref{fig:PhaseExploration}~\textbf{a} illustrates the procedure: starting from an initial set of parameters $\theta$ (a point in the phase diagram), we optimize $\theta$ by computing the gradients with QEP to find the (potentially local) maximum of the expectation value of interest.
    The only difference in the QEP procedure here is that the derivative of the simple loss function $\mathcal{L}=-y$ w.r.t. the output variable yields a trivial error signal of $-1$ which can then be inserted in the gradient calculation above according to Eq.~\eqref{eq:QEPDerivative}.

    Here, we exemplify the procedure by examining a slice through the phase diagram of the cluster Ising Hamiltonian~\eqref{eq:ClusterIsing} when we fix $g_{ZXZ}=-0.5\equiv\mathrm{const.}$ We optimize the N\'eel order parameter $y\equiv\langle {\hat X}_0 {\hat X}_4\rangle$, such that the loss function is simply $\mathcal{L}(y)\equiv - \langle {\hat X}_0 {\hat X}_4\rangle$. We show two example trajectories in Fig.~\ref{fig:PhaseExploration}~\textbf{b}. In both cases, the trajectories quickly converge to the $g_{ZZ}=0$ line and then move along it, going into the paramagnetic phase. Accordingly, the loss function, Fig.~\ref{fig:PhaseExploration}~\textbf{c}, rapidly decreases, with the decrease becoming slower as the trajectory moves along the $g_{ZZ}=0$ line.

    We suggest that, in general, this approach could be an efficient technique for exploring higher dimensional phase diagrams which cannot simply be mapped out without considerable effort by simply sweeping all of the parameters.

    \paragraph{Sensitivity optimization:}
    In the second unsupervised learning application, the aim is to maximize the \textit{derivative} of some expectation value of interest w.r.t. a certain parameter $\theta_j$. As Fig.~\ref{fig:PhaseExploration}~\textbf{d} illustrates, in contrast to the previous examples, we now start with two points, $\theta^{(1)}$ and $\theta^{(2)}$, in the phase diagram at which we compute expectation values $y_{1,2}\equiv \langle \hat A_\ell\rangle\rvert_{\theta^{(1,2)}}$ and maximize the slope calculated from the difference quotient of the output expectation values at these two points.
    Concretely, the corresponding loss function has the form
    \begin{align}\label{eq:lossSensitiviyOptimisation}
        \mathcal{L}(y_1, y_2) = -\left\lvert\frac{\langle \hat A_\ell \rangle\rvert_{\theta^{(1)}} - \langle \hat A_\ell \rangle\rvert_{\theta^{(2)}}}{\theta_j^{(1)} - \theta_j^{(2)}}\right\rvert.
    \end{align}
    An appealing application of this procedure could be to find the optimal working point of sensors, such as magnetic field sensors.
    
    To illustrate this scheme, we again consider the cluster Ising Hamiltonian~\eqref{eq:ClusterIsing} and optimize the slope of $\langle {\hat X}_0 {\hat X}_4\rangle$ w.r.t. the parameter $g_X$, which is proportional to the magnetic field.
    Fig.~\ref{fig:PhaseExploration}~\textbf{e} shows two sets of example trajectories and the corresponding loss functions, Fig.~\ref{fig:PhaseExploration}~\textbf{f}.  We observe that the trajectories converge to the phase boundary, where the slope is largest. For the first run, Fig.~\ref{fig:PhaseExploration}~\textbf{g}, we see that the trajectory moves along the phase boundary, since the slope is larger for smaller $g_{ZZ}$.
    We suggest that, in the future, this feature could be exploited to more generally map out phase boundaries.
      
\section{Conclusion and Outlook}

We have exploited Onsager reciprocity to derive a quantum version of equilibrium propagation, which in its classical form is one of the major general training techniques for neuromorphic platforms. We have shown that this can be used successfully even for situations where the input is effectively a quantum state (as in classifying quantum phases via supervised learning), as well as for unsupervised learning tasks related to exploring the phase diagrams of quantum simulators. In all of these cases, QEP can be applied even when classically, the Hamiltonian is hard or impossible to simulate.
A large variety of experimental platforms should be amenable to implementations of quantum equilibrium propagation.

During the final stage of completion of this manuscript and shortly before submission, two strongly related preprints appeared on the arXiv~\cite{massar2024equilibrium,scellier2024quantum}, also introducing a quantum version of equilibrium propagation, with somewhat different use cases.

%

\appendix

\section*{Appendix}

\subsection*{Appendix: Some details for the supervised learning example }

We considered a cluster Ising chain of $N=8$ spins, and a sensor of $2$ qubits, such that the total Hilbert space is $1024$-dimensional. This allows efficient exact diagonalization using the Lanczos algorithm applied to sparse matrices, to find the ground state of the coupled system. The three considered output operators are projectors onto three states, each with a definite combination of the sensor operators ${\hat Z}_{1'}$ and ${\hat Z}_{2'}$, as shown in the figure; e.g. ${\hat P}_{1,-1}=(1 + {\hat Z}_{1'})(1-{\hat Z}_{2'})/4$  projects onto the combination $(+1,-1)$ and would be used to indicate the ferromagnetic phase, which is reached when $g_{\rm ZZ}$ dominates. The expectation value is correspondingly the probability to observe this particular combination in a projective measurement of these two operators. The training samples are drawn uniformly from the triangular phase diagram shown in the figure, where (following convention in the literature) we set the sum of all three coupling parameters to $4$. The phase boundaries are known for this benchmark quantum many-body model~\cite{liu2023model}, which allows us to provide the correct labels: in each phase, one of the projectors is $1$, while the other two are $0$ (``one-hot-encoding''), and the assignment of the three combinations of $(Z_{1'},Z_{2'})$ to the three phases is defined in a fixed arbitrary way.  

Gradients are obtained using the symmetric nudging procedure, by adding the nudged output operators $\sum_j \nu_\ell {\hat A}_\ell$ to the Hamiltonian, where in our case ${\hat A}_\ell={\hat P}_\ell$ and $l\in {(+1,+1),(+1,-1),(-1,-1)}$. The ground state for the nudged Hamiltonian (for both signs of nudging) is re-calculated using sparse Lanczos diagonalization. Interestingly, our experiments have shown that this is numerically more efficient than attempting to obtain the exact linear response using first-order perturbation theory, which involves solving a linear system of equations when applying $(E-{\hat H})^{-1}$ to the perturbed ground state. After the approximate gradient has been obtained using QEP, we use it inside an Adam adaptive gradient descent optimizer to update the parameters---which would be possible also for the QEP gradients obtained in real experiments. The learning rate employed in the numerical examples was set to $0.01$.

Test accuracies are measured on a test set of $200$ points that are also uniformly randomly distributed across the triangular space and which are fixed before the training run. We distinguish two measures of accuracy: In the ``many queries'' version, we imagine that one would obtain the expectation values of the three projectors, which requires multiple shots, and assign as official outcome the phase whose associated projector has the largest expectation value (largest measurement probability). This is similar to how accuracy would be assessed for classical machine-learning classification models. In the ``single shot'' version, we imagine to run inference only once (equilibrating to the ground state once) and performing a single measurement of the two sensor operators. The outcome will be declared correct if the combination matches the correct label of the true underlying phase for this parameter combination. Single-shot performance is more difficult, but even so the training results show that single-shot accuracy can also reach relatively high values. Repetition, e.g. using three shots and taking a majority vote when possible, will quickly boost the accuracy (until it reaches the 'many queries' result in the limit of many shots).

For assessing the influence of shot noise, we replace the exact expectation values by numerical values that are drawn from a Gaussian distribution centered around that value, with the correct variance ${\rm Var}{\hat A_\ell}/M$, where $M$ is the number of measurement shots (recall each of those will usually require a new equilibration, unless one adopts some of the strategies mentioned in the main text). Replacing the true distribution by a Gaussian is a reasonable approximation unless $M$ is very small.

\subsection*{Appendix: Different approaches for nudging }

In classical EP, the idea is to add the loss function $\mathcal{L}(y,y^{\rm target}(x))$ to the energy, multiplied by $\beta$, where $\beta$ is a small constant. Instead, in the main text we advocated simply adding $\sum_\ell \nu_\ell {\hat A}_\ell$ to the Hamiltonian, where ${\hat A}_\ell$ are the output operators and $\nu_\ell=\beta \partial {\mathcal L} / \partial y_\ell$ is the nudge force, since for small $\beta$  this produces the force needed to elicit the linear response required for the gradient. If we were instead to translate directly the classical EP prescription to the quantum Hamiltonian, we could use $\beta {\mathcal L}({\hat y}, y^{\rm target}(x))$, with the operator version of the outputs,  ${\hat y}_\ell={\hat A}_\ell$ for $\ell \in {\mathcal S}_{\rm out}$, replacing the expectation values $y_\ell$. Expanding this to linear order in ${\hat A}_\ell$,  we would get the same result as our ansatz. The higher-order terms in $\beta \mathcal{L}$ would lead to further corrections to the Hamiltonian, and depending on the quantum fluctuations in ${\hat A}_\ell$ these might be as large as the linear-order term itself (e.g. for mean-squared error loss functions, these could correspond to a stiffening of the potential acting on ${\hat A}_\ell$). Since these higher-order terms might also be more difficult to implement experimentally, depending on the shape of ${\hat A}_\ell$, and since they are, at the same time, not needed to evaluate the gradient, we opted for the procedure explained in the main text, adding only linear terms to the Hamiltonian. It would be interesting in the future to compare the various approaches for finite nudging, where $\beta$ is not small (which is actually the case for the numerical experiments).

\subsection*{Appendix: Further details for unsupervised learning examples}

\subsubsection*{Phase exploration}
We again consider a cluster Ising Hamiltonian, of $10$ spins, and search for the points in the phase diagram that maximize $\langle \hat X_0 \hat X_4\rangle$, hence, $\hat X_0 \hat X_4$ is the output operator. This is expected to be maximized in the paramagnetic phase.
Initial parameters will typically be chosen randomly. For illustrative purposes, we manually select the initial parameters, i.e. the starting points of the trajectories in Fig.~\ref{fig:PhaseExploration}~\textbf{b}.
Concretely, the two starting points are $g_X = -0.1$, $g_{ZZ} = 0.4$ (run 1) and $g_X = 0.9$, $g_{ZZ} = 0.9$ (run 2).
We obtain gradients according to the QEP procedure by switching on the coupling to the output operator $\nu \hat X_0 \hat X_4$. As in the supervised learning example, the ground state is computed using sparse Lanczos diagonalization.
For the results shown in Fig.~\ref{fig:PhaseExploration}~\textbf{b}, both the learning rate and the nudge parameter $\nu$ were set to $0.1$ 

\subsubsection*{Sensitivity optimization}
In the second unsupervised learning example, we search for the largest slope of $\langle \hat X_0 \hat X_4\rangle$ w.r.t. $g_X$ in the phase diagram of the cluster Ising Hamiltonian~\eqref{eq:ClusterIsing} with $10$ spins by optimizing the loss function~\eqref{eq:lossSensitiviyOptimisation}.
Concretely, we consider two different values of $g_X^{(1,2)}$ while the other parameters are the same ($g_{ZZ}$ is trained while $g_{ZXZ}=-0.5\equiv\mathrm{const.}$).

During one step of the training, we update $g_X^{(1)}$, $g_X^{(2)}$ and $g_{ZZ}$.
To obtain the necessary gradients, we need to compute the following derivatives (for brevity we denote $\partial/\partial \theta_j$ by $\partial_{\theta_j}$)
\begin{align}
    \partial_{g_X^{(1)}}\mathcal{L}
    = & \varepsilon \, \frac{\partial_{g_X^{(1)}}\langle \hat X_0 \hat X_4 \rangle\rvert_{g_X^{(1)}}}
    {\lvert g_x^{(1)} - g_x^{(2)}\rvert} \notag \\
    & +
    \mathrm{sgn}\,(g_X^{(1)} - g_X^{(2)}) \frac{\mathcal{L}}{\lvert g_X^{(1)} - g_X^{(2)}\rvert}
    \\
    \partial_{g_X^{(2)}}\mathcal{L}
    = & - \varepsilon \, \frac{\partial_{g_X^{(2)}}\langle \hat X_0 \hat X_4 \rangle\rvert_{g_X^{(2)}}}
    {\lvert g_x^{(1)} - g_x^{(2)}\rvert} \notag \\
    & -
    \mathrm{sgn}\,(g_X^{(1)} - g_X^{(2)}) \frac{\mathcal{L}}{\lvert g_X^{(1)} - g_X^{(2)}\rvert} \\
    \partial_{g_{ZZ}}\mathcal{L}
    = & \varepsilon \, \partial_{g_{ZZ}}
    \frac{\left(\langle \hat X_0 \hat X_4 \rangle\rvert_{g_X^{(1)}} - \langle \hat X_0 \hat X_4 \rangle\rvert_{g_X^{(2)}}\right)}
    {\lvert g_x^{(1)} - g_x^{(2)}\rvert}
\end{align}
with $\varepsilon\equiv \mathrm{sgn}\,[\langle \hat X_0 \hat X_4\rangle \vert_{g_X^{(1)}} - \langle \hat X_0 \hat X_4\rangle \vert_{g_X^{(2)}}]$.
In all of the expressions, we use QEP to extract
$\partial_{\theta_j} \langle \hat X_0 \hat X_4\rangle\rvert_{g_X^{(1,2)}}$ with $\theta_j\in\{g_X^{(1)}, g_X^{(2)}, g_{ZZ}\}$.

To that end, as outlined in the main text, we approximate the derivative by comparing the nudged and the free expectation value
\begin{align}
    \frac{\partial}{\partial \theta_j} \langle \hat X_0 \hat X_4 \rangle
    &
    \approx \frac{\langle \hat X_0 \hat X_4\rangle\big\rvert_{\nu = \beta \varepsilon} - \langle \hat X_0 \hat X_4\rangle\big\rvert_{\nu = 0}}{\beta}.
\end{align}
As before, to evaluate the expectation value in the nudged phase, we couple to the output operator by adding $\nu \hat X_0 \hat X_4$ to the Hamiltonian. Ground states in the free and the nudged phase are again computed using sparse Lanczos diagonalization.

To illustrate the procedure, we consider two different starting points in the phase diagram: $g_X^{(1)} = -0.2$, $g_X^{(2)} = -1.5$, $g_{ZZ} = -1.5$ (run 1; each points starts in a different phase) and $g_X^{(1)} = -0.5$, $g_X^{(2)} = 0.3$, $g_{ZZ} = 1.0$ (run 2; both points are in the same phase).
Both the learning rate and the nudging parameter were set to $0.1$.

\end{document}